\documentclass[sigconf]{acmart}

\copyrightyear{2022} 
\acmYear{2022} 
\setcopyright{acmlicensed}\acmConference[WWW '22]{Proceedings of the ACM
Web Conference 2022}{April 25--29, 2022}{Virtual Event, Lyon, France}
\acmBooktitle{Proceedings of the ACM Web Conference 2022 (WWW '22), April
25--29, 2022, Virtual Event, Lyon, France}
\acmPrice{15.00}
\acmDOI{10.1145/3485447.3512110}
\acmISBN{978-1-4503-9096-5/22/04}

\usepackage[labelfont=bf]{caption}
\usepackage{subcaption}
\usepackage{booktabs, array, multirow}
\usepackage[algoruled]{algorithm2e}
\usepackage{algorithmic}
\usepackage{listings}
\usepackage{fancyvrb}
\fvset{fontsize=\small}

\AtBeginDocument{%
  \providecommand\BibTeX{{%
    \normalfont B\kern-0.5em{\scshape i\kern-0.25em b}\kern-0.8em\TeX}}}
    
\copyrightyear{2021} 
\acmYear{2021} 

\begin{document}
\title{Mutually-Regularized Dual Collaborative Variational Auto-encoder for Recommendation Systems}

\author{Yaochen Zhu}
\affiliation{%
  \institution{School of Remote Sensing and Information Engineering, Wuhan University}
  \city{Wuhan}
  \country{China}
}
\affiliation{%
  \institution{Media Lab, Tencent}
  \city{Shenzhen}
  \country{China}
}

\author{Zhenzhong Chen}
\authornote{Corresponding Author: Zhenzhong Chen.}
\affiliation{%
  \institution{School of Remote Sensing and Information Engineering, Wuhan University}
  \city{Wuhan}
  \country{China}
}
\email{zzchen@whu.edu.cn}

\begin{abstract}
Recently, user-oriented auto-encoders (UAEs) have been widely used in recommender systems to learn semantic representations of users based on their historical ratings. However, since latent item variables are not modeled in UAE, it is difficult to utilize the widely available item content information when ratings are sparse. In addition, whenever new items arrive, we need to wait for collecting rating data for these items and retrain the UAE from scratch, which is inefficient in practice. Aiming to address the above two problems simultaneously, we propose a mutually-regularized dual collaborative variational auto-encoder (MD-CVAE) for recommendation. First, by replacing randomly initialized last layer weights of the vanilla UAE with stacked latent item embeddings, MD-CVAE integrates two heterogeneous information sources, \textit{i.e.}, item content and user ratings, into the same principled variational framework where the weights of UAE are regularized by item content such that convergence to a non-optima due to data sparsity can be avoided. In addition, the regularization is mutual in that user ratings can also help the dual item content module learn more recommendation-oriented item content embeddings. Finally, we propose a symmetric inference strategy for MD-CVAE where the first layer weights of the UAE encoder are tied to the latent item embeddings of the UAE decoder. Through this strategy, no retraining is required to recommend newly introduced items. Empirical studies show the effectiveness of MD-CVAE in both normal and cold-start scenarios. Codes are available at \url{https://github.com/yaochenzhu/MD-CVAE}.

\end{abstract}



\acmSubmissionID{1952}

\maketitle

\section{Introduction}

Personalized recommendation plays a pivotal role in modern web services. Collaborative filtering (CF), which recommends new items by exploiting similarity patterns in users' historical interactions, has become a fundamental component of existing recommender systems \cite{koren2015advances, zhang2016collaborative}. However, the performance of CF degenerates significantly when collected historical ratings are sparse. In addition, it cannot recommend items that have yet received any ratings from users. Consequently, hybrid methods that utilize item content as auxiliary information to address the sparsity and cold-start problems have gained more attention among researchers \cite{strub2016hybrid, dong2017hybrid}. 

\begin{figure}
\centering
\includegraphics[width=0.8\linewidth]{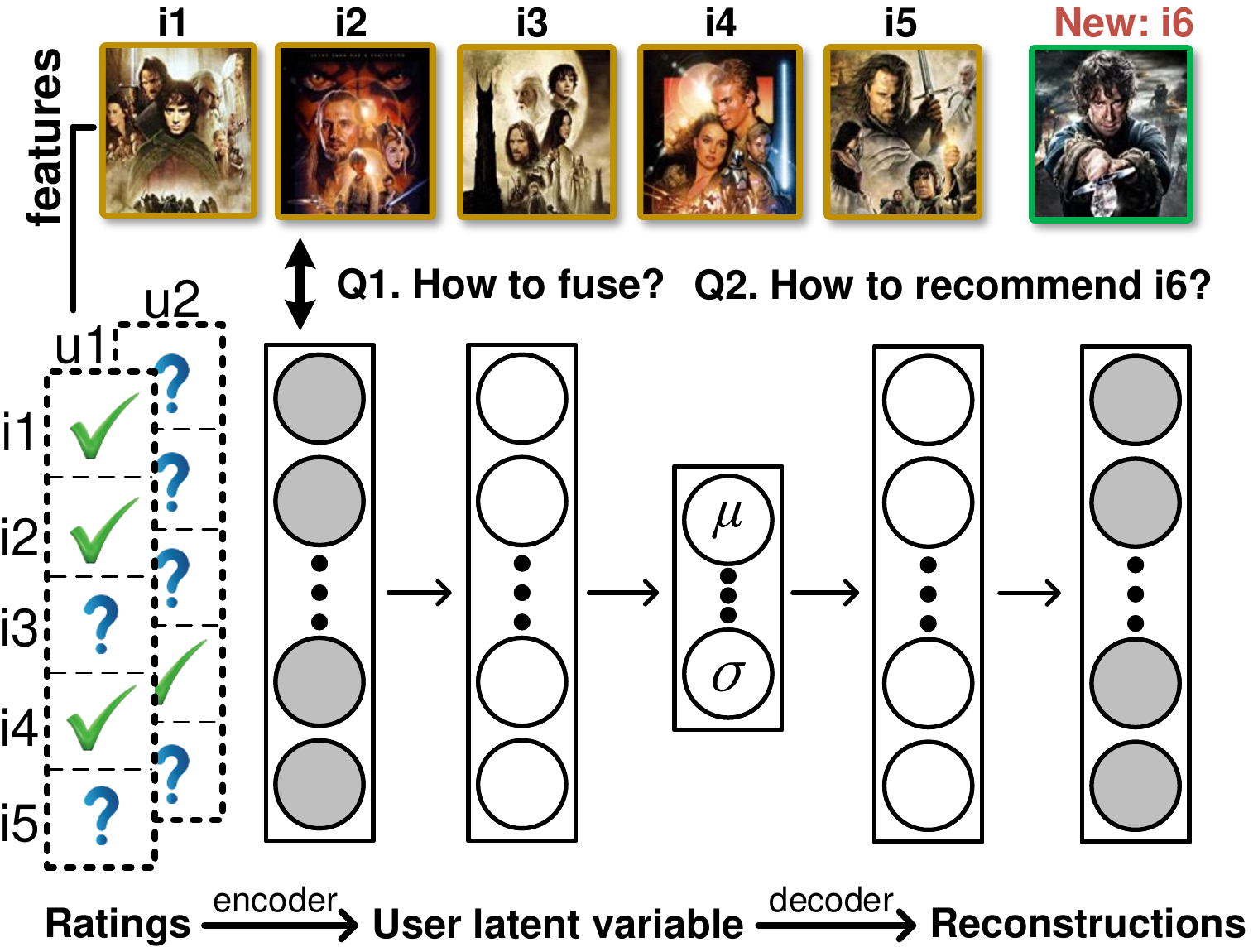}
\caption{An overview of UAE-based recommenders and two key problems. Q1. How to utilize item content information where no item latent variable is explicitly modeled? Q2. How to enable UAE to recommend cold-start items like i6?}
\label{fig:teaser}
\end{figure}

Item content can be naturally incorporated into traditional matrix factorization (MF)-based CF methods \cite{burke2002hybrid}. These methods model users' preference to items by the inner product between the corresponding user, item embeddings, where item content information can be directly used to constrain the factorized item collaborative embeddings to prevent the model from stuck into a bad solution due to the sparsity of ratings \cite{li2017collaborative}. However, since these factorization-based methods can only capture the linear similarity between users and items, their model capacity is severely restricted for the recommendation demand of large-scale modern online platforms.

Recently, auto-encoders (AEs) have been widely adopted in CF \cite{sedhain2015autorec,yi2021cross,zhu2022deep} due to their ability to learn compact user representations from sparse inputs and to encode non-linear user similarities based on their historical interactions. As a Bayesian version of AE, variational auto-encoder (VAE) demonstrates superiority because it explicitly models the uncertainty in user latent variable as the embedding variance and is more robust to rating noise  \cite{ma2019learning,shenbin2020recvae,xie2020multimodal}. However, since these AEs are user-oriented, \textit{i.e.}, they encode users' historical interactions into user latent embeddings and from them directly reconstruct ratings for recommendations, latent item factors are unintentionally eliminated from the model (see Fig. \ref{fig:teaser}). Although such a characteristic is efficient to fold-in (\textit{i.e.,} infer the latent user variable and reconstruct the ratings for) new users whose click histories are not present in the training data \cite{pang2019novel,wu2020hybrid, luo2020deep} (Folding-in new users for MF-based methods, in comparison, requires laborious iterative inference process), the direct way to incorporate item content information is obscured. This severely degenerates the recommendation performance for items with sparse ratings. Faced with this dilemma, several pioneering works have explored methods to utilize the widely available item content information in UAEs. An exemplar method is the collective VAE (CoVAE) \cite{chen2018collective}, where item feature co-occurrences are used to train UAE together with user ratings (which are essentially item co-purchases). Although improvement has been observed compared with the vanilla UAEs, since user ratings and item feature co-occurrences are two heterogeneous information sources, such a treatment lacks some theoretical interpretability. In addition, since item features are only used as extra training samples, the collaborative and content modules of CoVAE are loosely coupled, and to make out-of-matrix predictions, \textit{i.e.}, to recommend newly introduced items, is still infeasible. 

To address the above challenges, we propose a Bayesian generative model called mutually-regularized dual collaborative variational auto-encoder (MD-CVAE) for recommendations. By replacing randomly initialized last layer weights of the vanilla UAE with stacked latent item embeddings, MD-CVAE tightly couples UAE and an item content VAE into a unified variational framework where item content information can be introduced to regularize the weights of UAE, preventing it from converging to a non-optima when historical ratings are sparse. In turn, the collaborative information in user ratings constrains the item content VAE to learn more recommendation-oriented item representations. Furthermore, we propose a symmetric inference strategy for MD-CVAE that ties the first layer weights of UAE encoder to the latent item embeddings of UAE decoder. Through this mechanism, MD-CVAE can directly recommend new items without a time-consuming model retraining process, where latency can be substantially reduced. The main contribution of this paper can be summarized as:

\begin{itemize}
    \item By viewing UAE weights from an item embedding perspective, we propose a hierarchical Bayesian generative framework, MD-CVAE, that seamlessly unifies UAE with an item content representation learning module to address the sparsity and cold-start problems of vanilla UAEs.
    
    \item The MD-CVAE is tightly coupled in that item content embeddings regularize the weights of UAE to prevent bad model solutions due to rating sparsity, while user ratings in turn help the item content embedding module to learn more recommendation-oriented content representations.
    
    \item A symmetric inference strategy is proposed for MD-CVAE to address the out-of-matrix recommendation problem where existing UAE-based hybrid recommender systems fail. Experiments on real-world datasets show that MD-CVAE out-performs various state-of-the-art baselines.
\end{itemize}

\section{Problem Formulation}
\label{sec:prob}

In this paper, we consider recommendations with implicit feedback \cite{hu2008collaborative}. The user-item interactions are represented by an $I$ by $J$ binary rating matrix $\mathbf{R}$ where each row $\mathbf{r}_{i}^{T}$ denotes whether user $i$ has interacted with each of the existing $J$ items. The item content is denoted by a $J$ by $S$ matrix $\mathbf{X}$ where each row $\mathbf{x}_{j}^{T}$ is the extracted content feature for item $j$\footnote{\small{Subscript $i$ and $j$ would be omitted when no ambiguity exists. Capital non-boldface symbols such as $R$, $X$ are used to denote the corresponding random vectors of $\mathbf{r}$, $\mathbf{x}$, where exceptions such as $I$, $J$ are easily identified from contexts. $X^{s}$, $V^{s}$ are used to denote the random matrices for stacked $\mathbf{x}$, $\mathbf{v}$. Capital boldface symbols such as $\mathbf{R}$, $\mathbf{X}$, $\mathbf{V}$ are used to denote matrices.}}. Given partial observations of $\mathbf{R}$ and item content $\mathbf{X}$, the primary target of MD-CVAE is to predict the remaining ratings in $\mathbf{R}$ with the support of item content $\mathbf{X}$ such that new relevant items can be automatically recommended even if interactions for some items are extremely sparse. Moreover, if $J'$ new items arrive with features $\mathbf{X}'$ ($J'$ by $S$) and no user interactions, another goal of MD-CVAE is to immediately recommend these cold-start items without a laborious model retraining process.

\section{Methodology}

\subsection{Generative Process}

\begin{figure}
\centering
\includegraphics[width=0.83\linewidth]{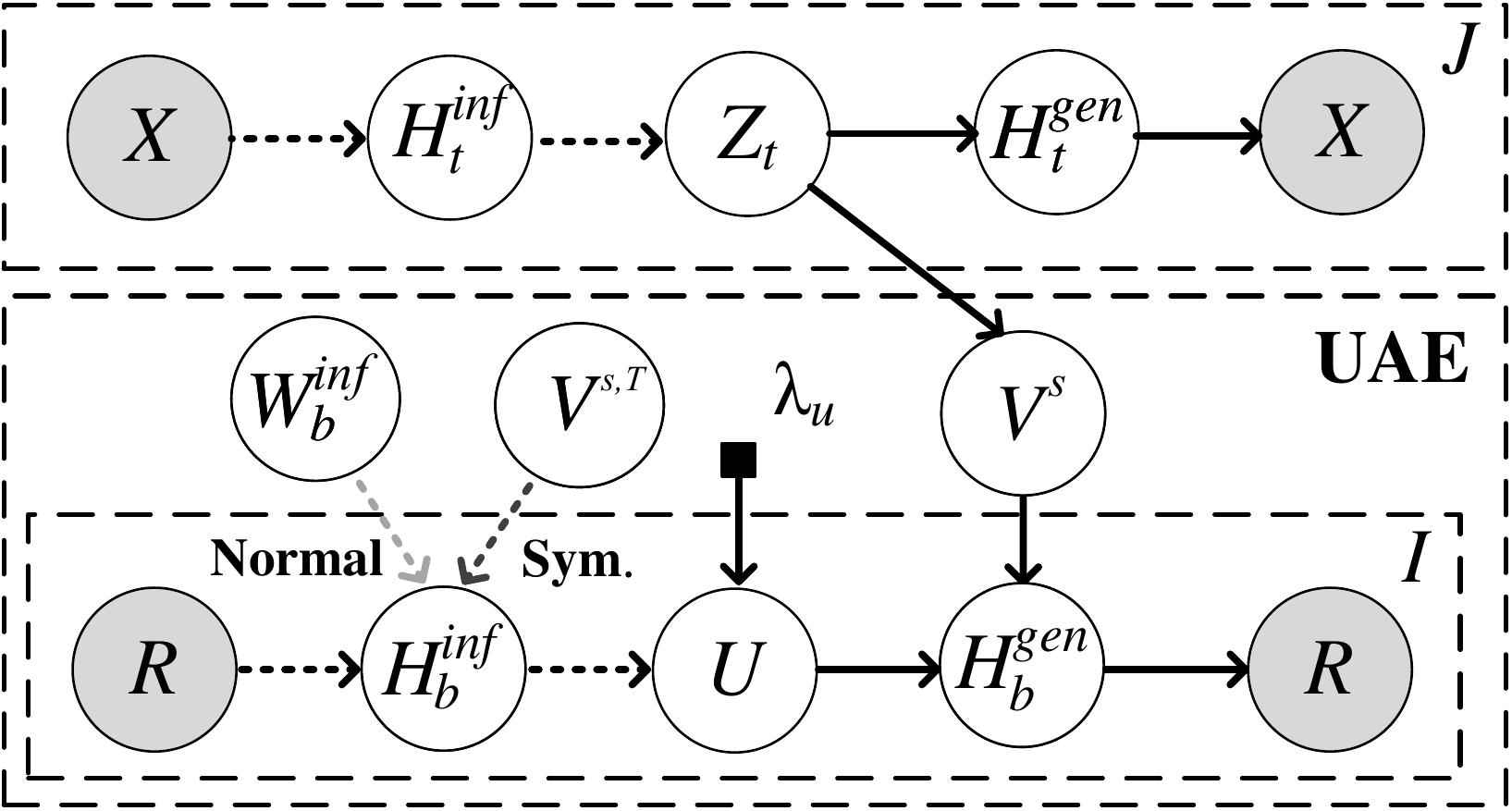}\caption{The probabilistic graphical model (PGM) of MD-CVAE. Solid lines denote the generative process whereas dashed lines denote the inference process. }
\label{fig:pgm}
\end{figure}

With a proper definition of the problems and goals, we are ready to introduce the mutually-regularized dual collaborative variational auto-encoder (MD-CVAE). The PGM of MD-CVAE is illustrated in Fig. \ref{fig:pgm}. In MD-CVAE, users and items are represented in two low-dimensional latent spaces of dimension $K_{u}$ and $K_{v}$, respectively. As with most probabilistic recommendation methods, the latent variable $\mathbf{u}$ for user $i$ is drawn from a Gaussian distribution as: 
\begin{equation}
    \mathbf{u} \sim \mathcal{N}(\mathbf{0}, \lambda _ {u} ^ {-1} \mathbf{I}_{K_{u}}).
\end{equation}
\noindent Different from the vanilla UAE (\textit{i.e.,} Multi-VAE in \cite{liang2018variational}) where the latent item factors are eliminated from consideration, we explicitly introduce $\mathbf{z}_{b}$ to embed the item collaborative information for item $j$, and draw it from another Gaussian distribution as:
\begin{equation}
    \mathbf{z}_{b} \sim \mathcal{N}(\mathbf{0}, \lambda _ {v} ^ {-1} \mathbf{I}_{K_{v}}).
\end{equation}
\noindent With the latent item collaborative embedding $\mathbf{z}_{b}$ explicitly introduced into the probabilistic graph, the widely available item content can be used to incorporate auxiliary information into the system. Specifically, MD-CVAE draws the latent item content variable from a standard normal distribution as follows:
\begin{equation}
    \mathbf{z}_{t} \sim \mathcal{N}(\mathbf{0}, \mathbf{I}_{K_{v}}).
\end{equation}
\noindent Importantly, to tightly couples the item collaborative and content variables, as with the collaborative variational auto-encoder (CVAE) \cite{li2017collaborative}, we set the latent item variable $\mathbf{v}$ to be composed of both item collaborative and content latent variables as follows:
\begin{equation}
\label{eq:fuse}
    \mathbf{v} = \mathbf{z}_{b} + \mathbf{z}_{t}.
\end{equation}
\noindent Given $\mathbf{z}_{t}$, then, the latent item variable $\mathbf{v}$ follows the conditional distribution $\mathcal{N}(\mathbf{z}_{t}, \lambda_{v}^{-1}\mathbf{I}_{K_{v}})$, which is the key to introduce mutual regularization between $\mathbf{v}$ and $\mathbf{z}_t$ in the MAP objective. We define the horizontally stacked latent item variables $\mathbf{v}$ for all $J$ existing items as $\mathbf{V}^{s} = [\mathbf{v}_{1}^{T}, \mathbf{v}_{2}^{T}, \cdots, \mathbf{v}_{J}^{T}]$ and the corresponding random matrix as $V^{s}$ (The superscript $s$ is short for stacked). To generate the rating vector  $\mathbf{r}$, we first embed the user latent variable $\mathbf{u}$ into a latent representation with an $N-1$-layer multi-layer perceptron $\mathrm{MLP} _ {u, gen}:\mathbb{R}^{K_{u}} \rightarrow \mathbb{R}^{K_{v}}$ with trainable weights $\boldsymbol{\theta}_{r}$ as $\mathbf{h} ^ {gen} _{b} (\mathbf{u}) = \mathrm{MLP} _ {u, gen} (\mathbf{u})$. Then, following Multi-VAE \cite{liang2018variational}, we draw the rating vector $\mathbf{r}$ from a multinomial distribution $p_{\boldsymbol{\theta}_{r}}(\mathbf{r} \mid \mathbf{u}, \mathbf{V}^{s})$ as follows:
\begin{equation}
\label{eq:multi}
    \mathbf{r} \sim Multi \Big(\mathrm{softmax}\big(\mathbf{V}^{s} \cdot \mathbf{h}^ {gen}_{b}(\mathbf{u})\big), \#Int\Big),
\end{equation}
where $\#Int$ is the number of interacted items for user $i$. Eq. (\ref{eq:multi}) defines an $N$-layer MLP on top of $\mathrm{MLP} _ {u, gen}$ where the stacked item latent variable $\mathbf{V}^{s}$ can be viewed as its last layer weights. Softmax is used as the activation to make the outputs valid multinomial parameters where the probability mass is summed to one. This MLP is similar to the decoder of the vanilla UAE, except that the last layer weights are replaced with latent item embeddings where item content information can be incorporated as Eq. (\ref{eq:fuse}). Through this mechanism, MD-CVAE generalizes the generative process of vanilla UAEs such that item side information can be tightly coupled with their decoder's last layer weights for hybrid recommendations. To emphasize that $\mathbf{V}^{s}$ also serves the last layer weights of the UAE decoder, we rewrite $p_{\boldsymbol{\theta}_{r}}(\mathbf{r} \mid \mathbf{u}, \mathbf{V}^{s})$ as $p_{\boldsymbol{\theta}_{r}, \mathbf{V}^{s}}(\mathbf{r} \mid \mathbf{u})$, where the union of $\boldsymbol{\theta}_{r}, \mathbf{V}^{s}$ represents the set of all decoder weights. Moreover, compared to CVAE that uses a linear MF as the collaborative backbone, MD-CVAE inherits the computational efficiency of UAEs to fold-in new users  whose interactive history is not included in the training set, where the whole ratings of a user can be generated from the user latent variables via $p_{\boldsymbol{\theta}_{r}, \mathbf{V}^{s}}(\mathbf{r} \mid \mathbf{u})$ by one forward pass.

The item content $\mathbf{x}$ is generated from $\mathbf{z}_{t}$ through $p(\mathbf{x} \mid \mathbf{z}_{t})$ parameterized by another dual item MLP. Specifically, if $\mathbf{x}$ is real-valued, it can be generated from a normal distribution $\mathcal{N}(\operatorname{MLP}_{i, gen}(\mathbf{z}_{t}), \lambda_{x}^{-1}\mathbf{I}_{S})$ where the output of the MLP denotes its mean and $\lambda_{x}$ is the precision; or if $\mathbf{x}$ is binary, we squash the output by sigmoid function and draw $\mathbf{x}$ from the corresponding Bernoulli distribution. Accordingly, we denote the random matrices for horizontally stacked $\mathbf{z}_{t}$ and $\mathbf{x}$ as $Z^{s}_{t}$ and $X^{s}$, respectively. With the above generative process defined, MD-CVAE is represented by the joint distribution of observable and hidden variables for user $i$ and all $J$ existing items as follows:
\begin{equation}
\label{eq:generative}
\begin{aligned}
    p_{\boldsymbol{\theta}}(R,X^{s},U,V^{s},Z^{s}_{t}) = & p_{\boldsymbol{\theta}_{r}, V^{s}} (R \mid U) \cdot p_{\boldsymbol{\theta}_{x}}(X^{s} \mid Z^{s}_{t}) \cdot \\
    &p(V^{s} \mid Z^{s}_{t}) \cdot p(Z^{s}_{t}) \cdot p(U),
\end{aligned}
\end{equation}
\noindent where $\boldsymbol{\theta}_{x}$ denotes trainable weights of the generative neural network for the item content and $\boldsymbol{\theta} = \{\boldsymbol{\theta}_{r}, \boldsymbol{\theta}_{x}\}$. Moreover, $p_{\boldsymbol{\theta}_{x}}(X^{s} \mid Z^{s}_{t})$ and $p(V^{s} \mid Z^{s}_{t})$ can be factorized into the product of per item distributions $\Pi_{j} p_{\boldsymbol{\theta}_{x}}(X \mid Z_{t})$, $\Pi_{j} p(V \mid Z_{t})$, respectively, due to the assumption of marginal independence among $J$ items. 

\subsection{Normal Inference: Vanilla MD-CVAE}
    
Given Eq. ({\ref{eq:generative}}), however, the non-linearity of generative process precludes us from calculating the posterior distribution of the latent variables analytically, as computing the model evidence $\log p_{\boldsymbol{\theta}}(R,X^{s})$ requires integrating the latent space of $U, V^{s}, Z^{s}_{t}$, which is intractable. Therefore, we resort to variational inference \cite{blei2017variational}, where we introduce variational distribution $q_{\boldsymbol{\phi}}(U, V^{s}, Z^{s}_{t} \mid R, X^{s})$ parameterized by deep neural networks with trainable parameters $\boldsymbol{\phi}$ and in $q_{\boldsymbol{\phi}}$ find the distribution closest to the true but intractable posterior measured by KL divergence as an approximation. According to the conditional independence implied by the PGM of MD-CVAE, the joint posterior of all hidden variables can be factorized into the product of three compact parts as follows:
\begin{equation}
q_{\boldsymbol{\phi}}(U, V^{s}, Z^{s}_{t} \mid R, X^{s}) = q_{\boldsymbol{\phi}}(U \mid R) \cdot q_{\boldsymbol{\phi}}(Z^{s}_{t} \mid X^{s}) \cdot q(V^{s} \mid Z^{s}_{t}),
\end{equation}
where $q(V^{s} \mid Z^{s}_{t})$ is assumed to be a conditional Gaussian distribution with $Z^{s}_{t}$ as the mean and a pre-defined fixed value as the variance. According to the variational approximation theory \cite{jordan1999introduction}, minimizing the KL divergence is equivalent to maximization of the evidence lower bound (ELBO) as follows:
\begin{equation}
\label{eq:elbo}
\begin{aligned} 
\mathcal{L}=& \mathrm{E}_{q_{\boldsymbol{\phi}}}[\log p_{\boldsymbol{\theta}}(R, X^{s}, U, V^{s}, Z^{s}_{t})-\log q_{\boldsymbol{\phi}}(U, V^{s}, Z^{s}_{t} \mid R, X^{s})] \\=& \mathrm{E}_{q_{\boldsymbol{\phi}}}[\log p_{\boldsymbol{\theta}}(R \mid U)+\log p(V^{s} \mid Z^{s}_{t})+\log p_{\boldsymbol{\theta}}(X^{s} \mid Z^{s}_{t})] \\ &-\mathrm{KL}\left(q_{\boldsymbol{\phi}}(Z^{s}_{t} \mid X^{s}) \| p(Z^{s}_{t})\right) -\mathrm{KL}\left(q_{\boldsymbol{\phi}}(U \mid R) \| p(U)\right) + \mathcal{C}, \end{aligned}
\end{equation}
\noindent where the term $\mathcal{C}$ includes the entropy of $q(V^{s} \mid Z^{s}_{t})$, which is constant due to its fixed variance. The two parameterized terms in the factorized posterior, \textit{i.e.,} $q_{\boldsymbol{\phi}}(U \mid R)$, $q_{\boldsymbol{\phi}}(Z^{s}_{t} \mid X^{s})$, correspond to the encoder of the UAE and the encoder of the dual item VAE in MD-CVAE, respectively. The maximum of Eq. (\ref{eq:elbo}) is achieved if and only if the variational distribution exactly matches true posterior. 

\subsection{Symmetric Inference: MDsym-CVAE}

The generative process of MD-CVAE allows us to incorporate widely available item content information to generate ratings from user latent variables for recommendations. Given the historical ratings $\mathbf{r}$ for user $i$, however, exploiting item content to support the inference of $\mathbf{u}$ that governs the generation of $\mathbf{r}$ is still infeasible, which may lead to large uncertainty in the estimation if observed ratings $\mathbf{r}$ for the user is sparse. In this section, we propose a symmetric inference strategy for MD-CVAE to solve this problem.

\begin{figure}
\centering
\includegraphics[width=0.99\linewidth]{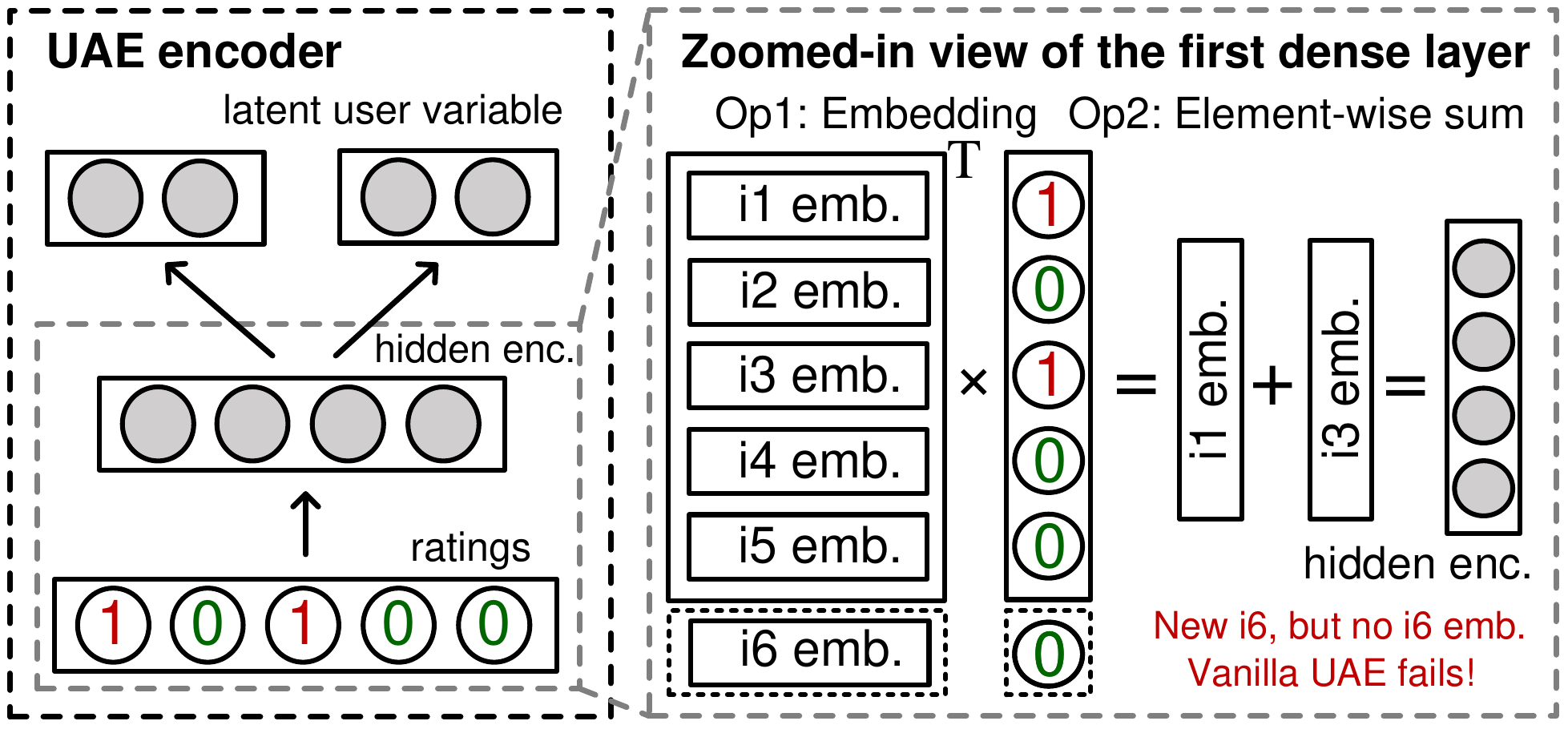}
\caption{The zoomed-in view of the first layer of UAE. Like the last layer of UAE decoder, the first layer of UAE encoder can also be viewed as latent item embeddings.}
\vspace{-5mm}
\label{fig:zoom}
\end{figure}

We demonstrate that, like the last layer weights of a UAE decoder, the first layer weights of a UAE encoder can also be interpreted as latent item embeddings. The reason is that, given that the inputs to UAEs are binary rating vectors denoting the implicit feedback, the first dense layer of the UAE encoder, which represents matrix-vector multiplication between its weights and a binary rating vector, can be decomposed into two basic operations, \textit{i.e.,} \textit{embedding} and \textit{element-wise sum} as follows (bias omitted for simplicity):
\begin{equation}
\label{eq:flw}
    \mathbf{h}_{b}^{inf} = \mathbf{W}^{inf}_{b} \cdot \mathbf{r} = \sum _ {j} \mathbb{I}(r_{j}=1) \cdot \mathbf{w}^{inf}_{b, j},
\end{equation}
where $\mathbb{I}$ is the indicator function and $\mathbf{w}^{inf}_{b, j}$ is the $j$th column of weight matrix $\mathbf{W}^{inf}_{b}$ (see Fig. \ref{fig:zoom}). Therefore, a one-to-one relationship can be established between the $j$th column of $\mathbf{W}^{inf}_{b}$ and the $j$th item, which allows us to view $\mathbf{W}^{inf}_{b}$ as vertically stacked embeddings of the $J$ existing items. Consequently, item content embeddings $\mathbf{z}_{t, j}$ can be fused with $\mathbf{w}^{inf}_{b, j}$ the same way as Eq. ($\ref{eq:fuse}$) to support the inference of $\mathbf{u}$. However, a better strategy is to reuse $\mathbf{v}_{j}$ in the UAE decoder as the weights $\mathbf{w}^{inf}_{b, j}$ by setting $\mathbf{W}^{inf}_{b} = \mathbf{V}^{s, T}$, where $T$ stands for matrix transposition, such that extra structure regularization is imposed on the UAE part of the MD-CVAE to avoid overfitting on sparse ratings. Since through this strategy, the UAE encoder in MD-CVAE is a symmetric version of the UAE decoder (by the first and the last layer), we use MDsym-CVAE to distinguish it from the vanilla MD-CVAE. Compared to MD-CVAE, MDsym-CVAE has another advantage of direct recommending cold-start items without a laborious model retraining process, which will be thoroughly discussed in the out-of-matrix prediction section. 

\subsection{Maximum A Posteriori Estimate}

Given Eq. (\ref{eq:elbo}), MD-CVAE can be trained end-to-end. However, a joint training may lead to the model's converging to an undesirable sub-optima where it relies solely on one information source for recommendations. Therefore, we adopt an EM-like optimization strategy as CVAE. First, for user $i$ we fix stacked item content random vectors, \textit{i.e.,} $Z^{s}_{t}$, to $\hat{Z}^{s}_{t}$, and maximize the following objective:
\begin{equation}
\label{eq:bstep}
\begin{aligned} 
\mathcal{L}^{MAP}_{b\_\mathrm{step}}&=\mathrm{E}_{q_{\boldsymbol{\phi}(U \mid R)}}\big[\log p_{\boldsymbol{\theta}}(R \mid U)  \big] - \frac{\lambda_{v}}{2} \cdot \|V^{s}-\hat{Z}^{s}_{t}\|^{2}_{F} \\ &  -\mathrm{KL}\left(q_{\boldsymbol{\phi}}(U \mid R) \| p(U)\right) - \frac{\lambda_{W}}{2} \cdot \sum_{l}\|\mathbf{W}_{b}^{(l)}\|_{F}^{2},
\end{aligned}
\end{equation}
\noindent where $\mathbf{W}_{b}^{(l)}$ is the $l$th layer weights of the UAE part of MD-CVAE (which are also assumed to be Gaussian), $\lambda_{W}$ is the precision, and $F$ is the Frobenius norm. Eq. (\ref{eq:bstep}) trains the UAE with an extra item content constraint to its last layer weights $V^{s}$. Then, for item $j$, we fix the $j$th row of the updated weights $V^{s}$ to $\hat{V}$ and optimize the following objective of the dual item content VAE as follows:
\begin{equation}
\label{eq:tstep}
\begin{aligned}
\mathcal{L}^{MAP}_{t\_\mathrm{step}}&=\mathrm{E}_{q_{\boldsymbol{\phi}(Z_{t} \mid X)}}\big[\log p_{\boldsymbol{\theta}}(X \mid Z_{t}) - \frac{\lambda_{v}}{2} \cdot \|\hat{V}-Z_{t}\|^{2}_{2} \ \big] \\ &-\mathrm{KL}\left(q_{\phi}(Z_{t} \mid X) \| p(Z_{t})\right) - \frac{\lambda_{W}}{2} \cdot \sum_{l}\|\mathbf{W}_{t}^{(l)}\|_{F}^{2},
\end{aligned}
\end{equation}
\noindent where $\mathbf{W}_{t}^{(l)}$ is the $l$th layer weights of the dual item content VAE of MD-CVAE. For both $\mathcal{L}^{MAP}_{t\_\mathrm{step}}$ and $\mathcal{L}^{MAP}_{b\_\mathrm{step}}$ the objective is composed of three parts: \textbf{(1)} The expected log-likelihood term, which encourages the latent user, item variables to best explain the observed user ratings and item content. \textbf{(2)} the MSE between the UAE weights and latent item content embeddings term, which tightly couples the user and item VAE where user ratings and item content mutually constrain each other to learn better representations for recommendations. \textbf{(3)} The KL with prior and weight decay terms, which impose regularization to the network weights to alleviate over-fitting. Liang \textit{et al.} \cite{liang2018variational} have shown that the KL term in the collaborative part could be too strong, which over-constrains the user collaborative embeddings. As a solution, they introduced a scalar $\beta$ to control the weight of the KL term in Eq. (\ref{eq:bstep}). Under such a setting, the model learns to encode as much information of $\mathbf{r}$ in $\mathbf{u}$ as it can at the initial training stages while gradually regularizing $\mathbf{u}$ by forcing it close to the prior as the training proceeds.

\subsection{Reparameterization Trick}
We use reparameterization trick to make the optimization of Eqs. (\ref{eq:bstep}) and (\ref{eq:tstep}) amenable \cite{kingma2014auto}. For the two Gaussian latent variables $\mathbf{z}_{g} \in \{\mathbf{u}, \mathbf{z}_{t}\}$,  we calculate their mean and logarithm of standard deviation via the corresponding encoder network and take samples $\mathbf{z}_{g}^{(l)}$ via the following differentiable bi-variate transformation:
\begin{equation}
\mathbf{z}_{g}^{(l)}   \Big([\boldsymbol{\mu} _ {\mathbf{z}_{g}}, \boldsymbol{\sigma}_{\mathbf{z}_{g}}], \boldsymbol{\epsilon}^{(l)}\Big)= \boldsymbol{\mu} _ {\mathbf{z}_{g}} + \boldsymbol{\epsilon}^{(l)} \odot \boldsymbol{\sigma}_{\mathbf{z}_{g}}, 
\end{equation}
\noindent where $\boldsymbol{\epsilon} ^{(l)} \sim \mathcal{N}(\mathbf{0}, \mathbf{I}_{K_{\{u,v\}}})$. The samples are then forwarded to the corresponding decoder network to reconstruct the inputs. When backward propagated, it was shown in  \cite{kingma2014auto} that the sampling gradient is an unbiased estimator of the gradient of $\mathcal{L}^{MAP}_{\{t,b\}\_\mathrm{step}}$ w.r.t. the network parameters. Moreover, previous work has shown that the variance of reparameterization trick is small enough such one sample for each data point suffices for convergence \cite{kingma2014auto}.

\subsection{In-Matrix Prediction}
\begin{figure}
\centering
\includegraphics[width=0.95\linewidth]{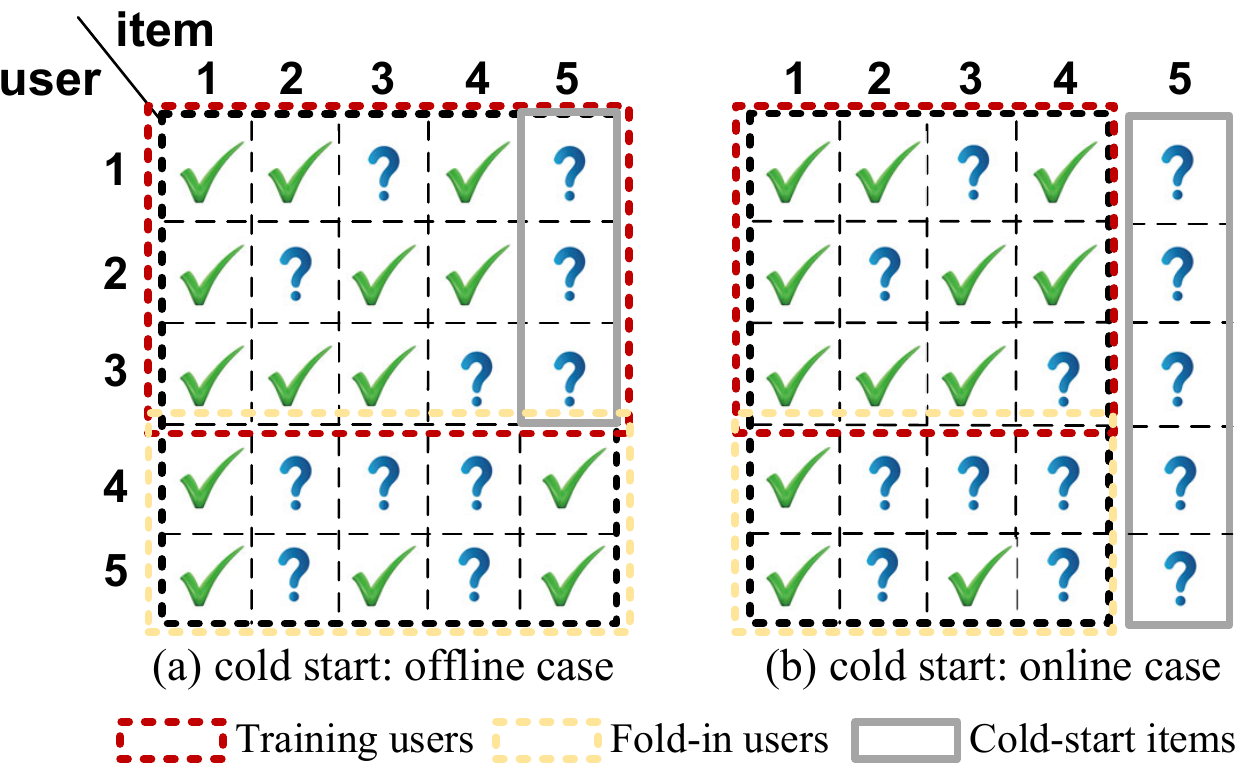}
\caption{Two cases of out-of-matrix prediction: (left) cold-start items already exist when the model is trained offline; (right) new items arrive in an online manner.}
\label{fig:cold}
\end{figure}
After the training of MD-CVAE, when conducting in-matrix predictions for a user, \textit{i.e.}, all candidate items have already been rated by at least one other user in the collected data, the historical interactions of that user can be used to calculate the mean of her latent embedding $\boldsymbol{\mu}_{\mathbf{u}}$ via the UAE encoder. Then, the multinomial probabilities of all hold-out items can be obtained via the UAE decoder, which are ranked to fetch $M$ most relevant ones for recommendations.

\begin{figure*}
\centering
\includegraphics[width=0.92\linewidth]{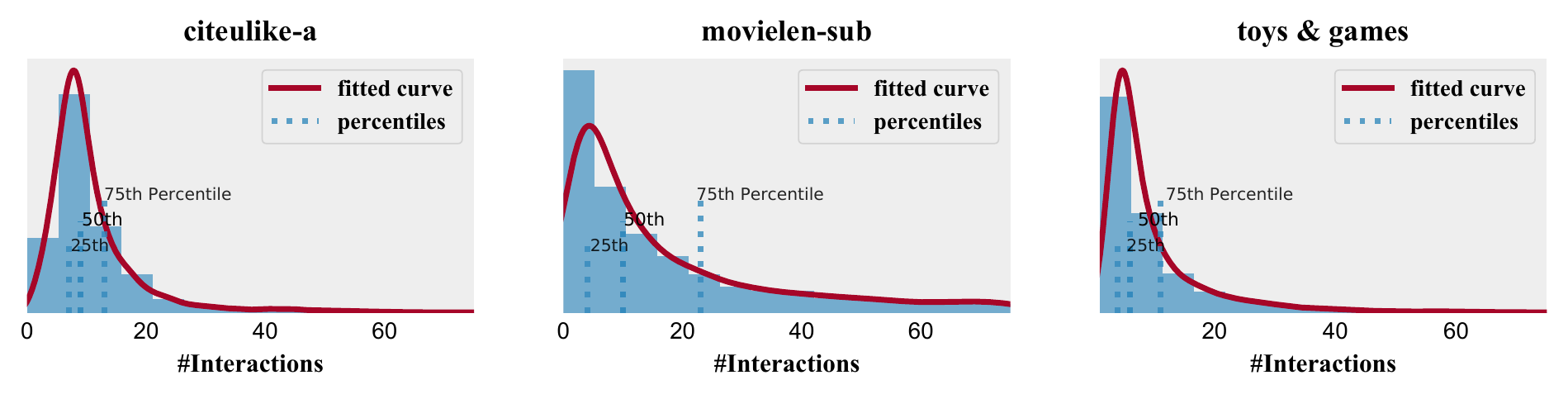}
\caption{The truncated item rating density distribution for \textit{citeulike-a}, \textit{movielen-sub}, and Amazon \textit{toys \& games} datasets. The red curves illustrate the estimated probability density functions and the light-blue dashed lines shows the percentiles. }
\label{fig:density}
\end{figure*}

\subsection{Out-of-Matrix Prediction}

The out-of-matrix prediction (\textit{i.e.} there exist items in the candidate set that have yet received any ratings from the users) could be categorized into the offline case and the online case, with differences between the two cases illustrated in Fig. \ref{fig:cold} for reference.

\subsubsection{The Offline Case} In the offline case, the never-before-visited items are mixed with normal items in the offline model training phase. Two problems could occur if such items exist for vanilla UAEs. First, in the rating generation process, since these items get no positive feedback, all embedded user latent variables $\mathbf{h}^ {gen}_{b}(\mathbf{u})$ are discouraged from being close to the last layer weights associated with these items in Eq. (\ref{eq:multi}). Therefore, these items would hardly get any recommendations, and users will be stuck to the same old items. In addition, in the model inference process, according to Eq. (\ref{eq:flw}), once initialized, the columns of the first encoder layer weights that correspond to cold-start items will never get updated. Therefore, if new users who have interacted with these items are folded-in for predictions, the inferred user latent variable $\boldsymbol{\mu}_{\mathbf{u}}$ would depend on randomly initialized weights, which would lead to unexpected behavior of the system. In contrast, the vanilla MD-CVAE addresses the first problem by constraining the last layer weights of UAE with item content embeddings, where recommendations of cold-start items can be made by utilizing similarity patterns between item content. Furthermore, MDsym-CVAE addresses the second problem by setting the first layer weights of UAE as the transpose of its content-regularized last layer weights, where the embeddings of fold-in users who have interacted with cold-start items will still contain content information of these items for recommendations.

\subsubsection{The Online Case} In the online case, $J'$ new items arrive when the model has already been deployed after training on rating data of existing $J$ items. Meanwhile, some users may have interacted with new items. When this occurs, UAE and vanilla MD-CVAE must wait for collecting users' interactions with these items and be retrained from scratch such that $J'$ new items can be properly recommended (A more direct reason is the dimensional mismatch between the network inputs/outputs and its first/last layer weights when $J'$ new items arrive). For MDsym-CVAE, since weights of these two layers are viewed as stacked item embeddings of existing $J$ items and are constrained to be close to corresponding item content embeddings, we can use item content embeddings inferred by the dual item content VAE as surrogates to the missing weights for $J'$ new items and conduct recommendations as if they were normal items.

If we denote the first and the last layer weights of the UAE part of MDsym-CVAE trained on existing $J$ items as $\mathbf{V}^{s}_{old} \in \mathbb{R}^{J \times K_{v}}$ and the horizontally stacked item content embeddings for $J'$ new items as $\mathbf{V}^{s}_{new} = [\boldsymbol{\mu}^{T}_{\mathbf{z}_{t}, J+1}, \boldsymbol{\mu}^{T}_{\mathbf{z}_{t}, J+2}, \cdots, \boldsymbol{\mu}^{T}_{\mathbf{z}_{t}, J+J'}] \in \mathbb{R}^{J' \times K_{v}}$, the new weights (or transposed weights) $\mathbf{V}^{s}_{surr} \in \mathbb{R}^{(J+J') \times K_{v}}$ are calculated as:
\begin{equation}
    \mathbf{V}^{s}_{surr} = [\mathbf{V}^{s}_{old} \ \| \ \mathbf{V}^{s}_{new}],
\end{equation}
where $\|$ denotes the operation of horizontal concatenation. $\mathbf{V}^{s}_{surr}$ can then be used as the surrogate for $\mathbf{V}^{s}$ in Eqs. (\ref{eq:multi}) and (\ref{eq:flw}) to infer user latent variable $\mathbf{u}$ and generate user rating vector $\mathbf{r}$. The remaining part of the network and the recommendation procedure of MD-CVAE remain unaltered. Through this strategy, predictions for cold-start items can be made as if they were normal items included in the training set, which reduces the frequency of model retraining and alleviates computational overhead. This is especially favorable in industrial applications where low latency matters.

\section{Experimental Analysis}

\subsection{Datasets}
\label{sec:dataset}

\begin{table}

\caption{Statistics of the datasets used in the paper}
\begin{tabular}{lcccc}
\toprule
dataset      & \#users & \#items & \%density &   \#features \\
\midrule
citeulike-a  & 5,551    & 16,980   & 0.217\%     & 8,000   \\
movielen-sub & 10,881   & 7,701    & 0.922\%     & 8,000   \\
toys \& games & 14,706    & 11,722   & 0.072\%   & 8,000 \\
\bottomrule
\end{tabular}
\label{tab:dataset}
\end{table}

We demonstrate the experimental results on three datasets. The first dataset, \textit{citeulike-a}, is a widely used hybrid recommendation benchmark collected by \cite{wang2011collaborative} for research article recommendations, where the title and abstract of each article are used as the item content. The second dataset, \textit{movielen-10m} \cite{harper2015movielens}, however, is rarely used to evaluate hybrid recommenders due to lack of item features. We collect the movie plots from IMDB and process the raw texts as \cite{li2017collaborative}. We keep movies with available plots and select a subset of users accordingly to form the \textit{movielen-sub} dataset. This dataset is included to evaluate MD-CVAE when collaborative information is comparatively sufficient. The last dataset is the Amazon \textit{toys \& games} dataset \cite{he2016ups}, which is larger in scale and sparser. We concatenate all reviews an item receives as the item features. For all three datasets, we obtain an 8,000-dimensional feature for each item from the item textual features as \cite{li2017collaborative}. In preprocessing, we randomly split the users by the ratio 8:1:1 for training, validation, testing, respectively. For each user, 20\% of interacted items are held out for evaluation. Table \ref{tab:dataset} summarizes the detail of the datasets after preprocessing.  Fig. \ref{fig:density} illustrates the distributions of interaction density for different items. The distribution of interaction density for all three datasets clearly demonstrates a long-tail characteristic, which hinders good recommendations for items with sparse ratings.

\subsection{Evaluation Metrics}

Two metrics are used to evaluate the model performance in this paper: Recall@$M$ and truncated normalized discounted cumulative gain (NDCG@$M$). For a user $i$, we first obtain the predicted rank of hold-out items by sorting their predicted multinomial probabilities. If we denote the item at rank $r$ by $j(r)$ and the set of hold-out items for user $i$ by $\mathcal{J}_{i}$, Recall@$M$ is then calculated as:
\begin{equation}
\operatorname{Recall} @ M(i)=\frac{\sum_{r=1}^{M} \mathbb{I}\left[j(r) \in \mathcal{J}_{i}\right]}{\min \left(M,\left|\mathcal{J}_{i}\right|\right)},
\end{equation}
\noindent where the denominator is the minimum of $M$ and the number of hold-out items. Recall@$M$ has a maximum of 1, achieved when all relevant items are ranked among top $M$ positions. Truncated discounted cumulative gain (DCG@$M$) is computed as
\begin{equation}
    \operatorname{DCG} @ M(i)=\sum_{r=1}^{M} \frac{2^{\mathbb{I}\left[j(r) \in \mathcal{J}_{i}\right]}-1}{\log (r+1)},
\end{equation}
\noindent which, instead of uniformly weighting all positions, introduces a logarithm discount function over the ranks where larger weights are applied to recommended items that appear at higher ranks. NDCG@$M$ is calculated by normalizing the DCG@$M$ to [0, 1] by the ideal DCG@$M$ where all relevant items are ranked at the top.

\subsection{Implementation Details}

Since the datasets that we include vary both in scale and scope, we select the architecture and hyperparameters of MD-CVAE on validation users through grid search. For the UAE part, we search networks with \{0,1,2\} hidden layer with hidden size \{50, 100, 150, 200, 250\} and keep the dual item VAE compatible with UAE, \textit{i.e.}, the dimension of latent item embedding equals to the hidden size of the last UAE layer. In addition, $\lambda_{v}$ is an important hyperparameter that controls the strength of mutual regularization in MD-CVAE, and for each architecture, we search $\lambda_{v}$ from \{0.1, 1, 2, 5, 10\}. We first layerwise pretrain the dual item content VAE and then iteratively train the UAE ($b$\_step) and the dual item VAE ($t$\_step) for 100 epochs. Adam \cite{kingma2014adam} is used as the optimizer with batches of 500 users. We randomly split the datasets into ten train/val/test splits. For each split, we keep the model with the best averaged value of Recall@20, Recall@40, and NDCG@100 on validation users. The metrics on test users averaged over ten splits are reported as model performance.

\subsection{Comparison with Baselines}

\begin{table*}[h]
\small
\centering
\caption{Comparisons between MD-CVAE and various baselines on citeulike-a, movielen-sub, and amazon datasets. The best method is highlighted in bold, whereas the best method in each part is marked with underline.}
\begin{tabular}{lccc|ccc|ccc}
\hline
         & \multicolumn{3}{c}{\textbf{citeulike-a}} & \multicolumn{3}{c}{\textbf{movielen-sub}}      & \multicolumn{3}{c}{\textbf{toys \& games}}    \\ \hline
         & Recall@20  & Recall@40  & NDCG@100 & Recall@20  & Recall@40  & NDCG@100 & Recall@20  & Recall@40  & NDCG@100 \\ \hline
MD-CVAE & \textbf{0.303} & \textbf{0.377} & \textbf{0.301} & \textbf{0.353} & \textbf{0.452}  & \textbf{0.381} & 0.141 & 0.188 & 0.102 \\
MDsym-CVAE & 0.295 & 0.374 & 0.297 & 0.347 & 0.449 & 0.377 & \textbf{0.147} & \textbf{0.191} & \textbf{0.106} \\ \hline
FM      & 0.231 & 0.312 & 0.238 & \underline{0.324} & 0.421 & \underline{0.357} & 0.088 & 0.121 & 0.062 \\
CTR      & 0.169 & 0.250 & 0.190 & 0.285 & 0.398 & 0.312 & 0.124 & 0.179 & 0.089 \\
CDL      & 0.209 & 0.295 & 0.226 & 0.311 & 0.405 & 0.339 & 0.133 & 0.181 & 0.092 \\
CVAE    & \underline{0.236} & \underline{0.334} & \underline{0.247} & 0.304 & \underline{0.422}  & 0.355 & \underline{0.139} & \underline{0.188} & \underline{0.094} \\ \hline
Multi-VAE & 0.269 & 0.346 & 0.274 & 0.326 & 0.423 & 0.357 & 0.114 & 0.157 & 0.082\\
CoVAE    & 0.247 & 0.338 & 0.260 & 0.338 & 0.436 & 0.367 & 0.120 & 0.174 & 0.085 \\
CondVAE  & 0.274 & 0.359 & 0.275 & \underline{0.341} & \underline{0.437} & 0.365 & \underline{0.132} & \underline{0.180} & \underline{0.094} \\ 
DICER & 0.279 & \underline{0.363} & 0.277 & 0.329 & 0.428 & 0.359 & 0.127 & 0.172 & 0.092\\
DAVE & \underline{0.281} & 0.362 & \underline{0.283} & 0.340 & 0.432 & \underline{0.371} & 0.125 & 0.177 & 0.086 \\

\hline
\end{tabular}
\label{tab:results}
\end{table*}

In this section, we compare MD-CVAE and MDsym-CVAE with state-of-the-art collaborative and hybrid recommendation baselines. Generally, for factorization-based methods, we search the optimal hyperparameters on validation users with Bayesian parameter search (BPS) \cite{dacrema2019we}, whereas for Multi-VAE-based baselines, we determine the optimal model structure with grid search.

\begin{itemize}
    \item \textbf{FM} (factorization machine) is a widely employed algorithm for hybrid recommendation that is robust to sparse inputs \cite{guo2017deepfm}. We use BPS as suggested in \cite{dacrema2019we} to find the optimal hyperparameters and loss function on validation users.
    
    \item \textbf{CTR} \cite{wang2011collaborative} learns the topics of item content via latent Dirichlet allocation (LDA) \cite{blei2003latent} and couples it with probabilistic matrix factorization (PMF) \cite{mnih2008probabilistic} for collaborative filtering. We find the optimal hyperparameters $a$, $b$, $\lambda_{u}$, $\lambda_{v}$ by BPS.
    
    \item \textbf{CDL} \cite{wang2015collaborative} replaces the LDA in CTR with a stacked Bayesian denoising auto-encoder (SDAE) \cite{vincent2010stacked} to learn the item content embeddings in an end-to-end manner. We set the mask rate of SDAE to 0.3 and search its architecture same as MD-CVAE.
    
    \item \textbf{CVAE} \cite{li2017collaborative} further improves over CDL by utilizing a VAE in place of the Bayesian SDAE, where a self-adaptive Gaussian noise is introduced to corrupt the latent item embeddings instead of corrupting the input features with zero masks. Grid search is used to find the optimal hyperparameters.
    
    \item \textbf{Multi-VAE} \cite{liang2018variational} breaks the linear collaborative modeling ability of PMF by designing a VAE with multinomial likelihood to model user ratings such that user collaborative information in ratings can be captured for recommendations.
    
    \item \textbf{CoVAE} \cite{chen2018collective} utilizes the non-linear Multi-VAE as the collaborative backbone and incorporates item feature information by treating feature co-occurrences as pseudo training samples to collectively train Multi-VAE with item features. 
    
    \item \textbf{CondVAE} \cite{pang2019novel} builds a user conditional VAE where user features are used as the condition. We extend the original CondVAE by replacing the categorical user features with averaged item features built from the interacted items, which we find have a better performance on all three datasets.
    
    \item \textbf{DICER} \cite{zhang2020content} is an item-oriented auto-encoder (IAE)-based recommender system where the item content information is utilized to learn disentangled item embeddings from their user ratings to achieve more robust recommendations.

    \item \textbf{DAVE} \cite{yi2021dual} hybrids an IAE with an adversarial UAE through neural collaborative filtering (NCF) \cite{he2017neural}. For a fair comparison, item content information is incorporated into the IAE part of DAVE by concatenation.
    
\end{itemize}

\subsubsection{In-matrix Case} We first evaluate the models for in-matrix predictions. The results are summarized in Table \ref{tab:results}. Table \ref{tab:results} comprises of three parts. The middle part shows the FM and three tightly-coupled hybrid baselines with linear collaborative modules. Generally, the performance improves with the increase of the representation learning ability of the item content embedding network. CVAE, which uses VAE to encode item content information, performs consistently better than CDL and CTR. However, simple methods such as FM can out-perform some deep learning baselines when their parameters are systematically searched, which is consistent with the findings in \cite{dacrema2019we}. The bottom part shows the vanilla UAE and hybrid baselines that employ UAE as the collaborative backbone. Since UAE can capture non-linear similarity among users, their performance improves over linear hybrid methods when datasets are comparatively dense. However, on the sparser \textit{toys \& games} dataset, they are out-performed by the shallow models. CoVAE does not always out-perform Multi-VAE, which indicates that the co-occurrence of item features does not necessarily coincide with the user co-purchases. Finally, DAVE performs almost consistently better than other UAE-based baselines, with an extra adversarial loss added to both user and item auto-encoders as regularization. 

However, the above UAE-based hybrid baselines are loosely-coupled, where user collaborative information cannot in turn guide the item content module to learn more recommendation-oriented content embeddings. Tightly coupling an item content VAE with UAE by introducing mutual regularization between UAE weights and item content embeddings, MD-CVAE out-performs the baselines on all three datasets. Specifically, we observe that MDsym-CVAE performs slightly worse than MD-CVAE on denser datasets (i.e., \textit{citeulike-a} and \textit{movielen-sub}). The reason could be that the symmetric structure in MDsym-CVAE trades some modeling capacity for the ability to recommend cold-start items. However, the better performance of MDsym-CVAE on \textit{toys \& games} dataset demonstrates that limiting the model capacity is not necessarily a disadvantage, as the extra structure regularization is conducive to preventing the model from overfitting when ratings are sparse.

\begin{figure}
\centering
\begin{subfigure}[b]{0.45\textwidth}
\includegraphics[width=\textwidth]{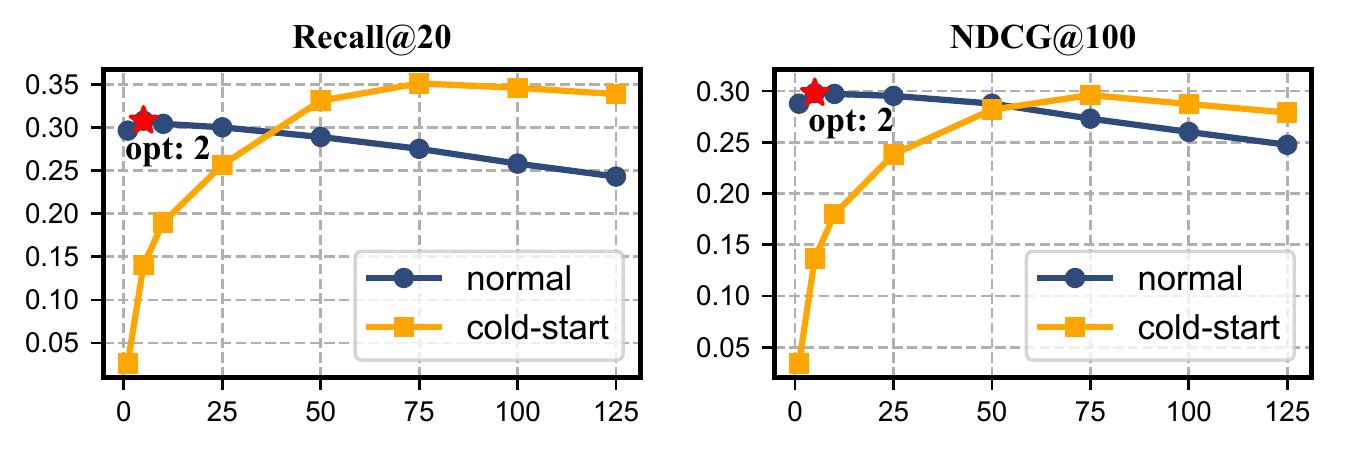}
\caption{\textit{citeulike-a} dataset}
\end{subfigure}

\begin{subfigure}[b]{0.45\textwidth}
\includegraphics[width=\textwidth]{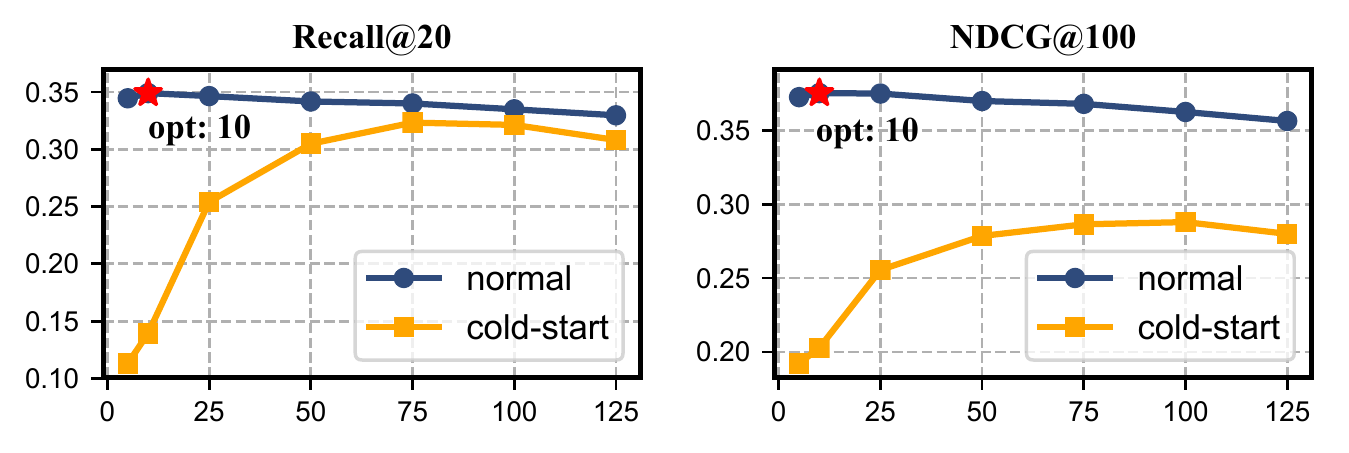}
\caption{\textit{movielen-sub} dataset}
\end{subfigure}

\begin{subfigure}[b]{0.45\textwidth}
\includegraphics[width=
\textwidth]{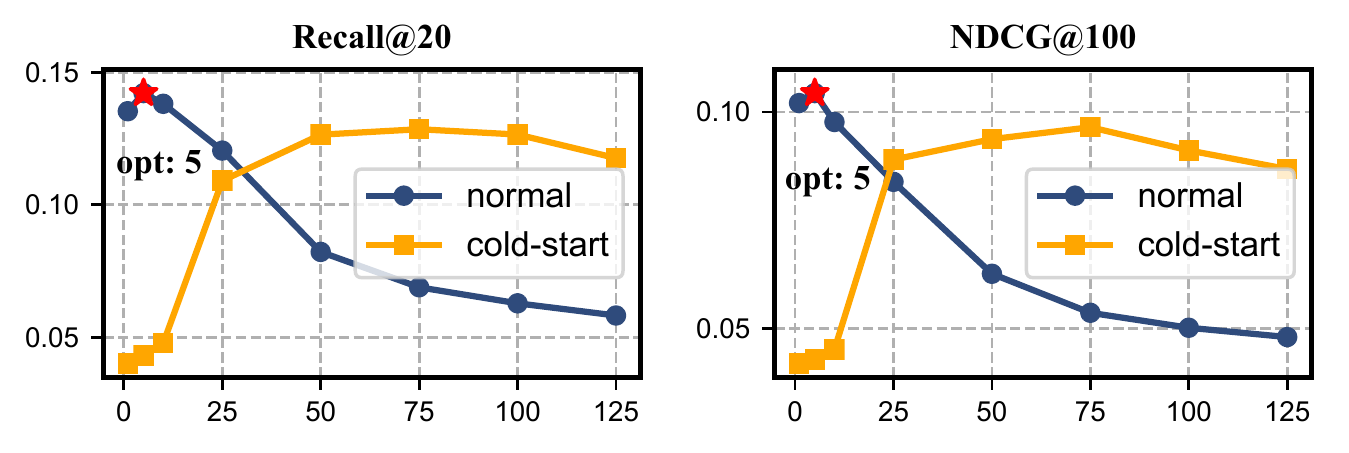}

\caption{\textit{toys \& games} dataset}
\end{subfigure}

\caption{Model performance with varied $\lambda_{v}$ on normal and cold-start items. {\color{red}$\star$} denotes the optimal $\lambda_{v}$ in in-matrix case.}

\vspace{-5mm}

\label{fig:wrt_lambda}
\end{figure}

\subsubsection{Out-of-matrix Case} 

\begin{table}
\caption{Comparisons between MDsym-CVAE and tightly coupled baselines (NI: Normal items; CI: Cold-start items).}
\begin{subtable}[t]{\columnwidth}
\centering
\caption{citeulike-a}\label{tab:a_cold}
\begin{tabular}{lcc}
\makebox[0.16\textwidth][c]{{}} & \makebox[0.31\textwidth][c]{Recall@20 \ (NI / CI)} & \makebox[0.31\textwidth][c]{NDCG@100 \ (NI / CI)} \\
\toprule 
MDsym-CVAE        & \textbf{0.290} \ /\ \ \textbf{0.251} & \textbf{0.286} \ / \ \textbf{0.249} \\
\midrule
CTR    & 0.169 \ /\ \ 0.209 & 0.189 \ /\ \ 0.207 \\
CDL    & 0.206 \ /\ \ 0.218 & 0.203 \ /\ \ 0.214 \\
CVAE   & \underline{0.238} \ /\ \ \underline{0.235} & \underline{0.236} \ /\ \ \underline{0.232} \\
\bottomrule
\end{tabular}
\end{subtable}

\bigskip

\begin{subtable}[t]{\columnwidth}
\centering
\caption{movielen-sub}\label{tab:m_cold}
\begin{tabular}{lcc}
\makebox[0.16\textwidth][c]{{}} & \makebox[0.31\textwidth][c]{Recall@20 \ (NI / CI)} & \makebox[0.31\textwidth][c]{NDCG@100 \ (NI / CI)} \\
\toprule 
MDsym-CVAE        & \textbf{0.351}\ /\ \ \textbf{0.309} & \textbf{0.369} \ /\ \ \textbf{0.275} \\
\midrule
CTR    & 0.284 \ /\ \ 0.164  & 0.310 \ /\ \ 0.195  \\
CDL    & 0.301 \ /\ \ 0.196  & 0.338 \ /\ \ 0.227  \\
CVAE   & \underline{0.318} \ /\ \ \underline{0.279} & \underline{0.342} \ /\ \ \underline{0.240} \\
\bottomrule
\end{tabular}
\end{subtable}

\bigskip

\begin{subtable}[t]{\columnwidth}
\centering
\caption{toys \& games}\label{tab:t_cold}
\begin{tabular}{lcc}
\makebox[0.16\textwidth][c]{{}} & \makebox[0.31\textwidth][c]{Recall@20 \ (NI / CI)} & \makebox[0.31\textwidth][c]{NDCG@100 \ (NI / CI)} \\
\toprule 
MDsym-CVAE        & \textbf{0.113} \ /\ \ \textbf{0.099} & \textbf{0.084} \ / \ \textbf{0.082} \\
\midrule
CTR    & 0.089 \ /\ \ 0.084 & 0.072 \ /\ \ 0.069 \\
CDL    & 0.096 \ /\ \ 0.090  & 0.077 \ /\ \ 0.079 \\
CVAE   & \underline{0.110} \ /\ \ \underline{0.094} & \underline{0.081} \ /\ \ \underline{0.080} \\
\bottomrule
\end{tabular}
\end{subtable}

\label{tab:results_cold}
\end{table}
Although CTR, CDL, CVAE are limited in their collaborative modeling ability, they are strong in out-of-matrix prediction where the UAE-based methods fail, because for CTR, CDL, CVAE, the inferred content embeddings of new items can be used in place of the latent item variables to calculate their inner-product with user latent variables for recommendations \cite{wang2011collaborative}. In this section, we compare the out-of-matrix performance of MDsym-CVAE with the three strong tightly-coupled baselines. In the experiment, we first fix the number of cold-start items, \textit{i.e.}, 1,600 for \textit{citeulike-a} (10\%), 160 for \textit{movielen-sub} (2\%), 1,200 (10\%) for \textit{toys \& games}, and evaluate the models separately on normal and cold-start items w.r.t varied $\lambda_{v}$. We then choose the optimal $\lambda_{v}$ and compare the results of MDsym-CVAE with the three strong baselines. Finally, we explore the influence of the number of cold-start items on recommendation performance for all tightly-coupled models. 

Fig. \ref{fig:wrt_lambda} shows the performance of MDsym-CVAE on normal and cold-start items with varied $\lambda_{v}$. In Fig. \ref{fig:wrt_lambda} we can find that generally, the recommendation performance for both normal and cold-start items improves first and then deteriorates with the increase of $\lambda_{v}$. Specifically, when $\lambda_{v}$ exceeds the in-matrix optimal value, which is denoted by red stars in Fig. \ref{fig:wrt_lambda}, both metrics drop monotonically for normal items, while they increase and then decrease for cold-start items. It is reasonable, since a larger $\lambda_{v}$ imposes a stronger constraint on UAE weights, which is beneficial for recommending cold-start items as their content embeddings are expected to resemble more to their corresponding weights if they would have appeared in the training set, while it may hurt predictions for normal items as it limits the model capacity when rating information is sufficient. We choose the optimal $\lambda_v$ for MDsym-CVAE where the performance on normal items drops no more than 80\% to keep the balance between predicting normal and cold-start items, which leads to $\lambda_v=50, 50, 10$ for the three datasets, respectively. The same criterion is used to select the optimal $\lambda_v$ for the three PMF-based tightly-coupled hybrid baselines. The performance comparisons are summarized in Table \ref{tab:results_cold}. From Table \ref{tab:results_cold} we can find that MDsym-CVAE performs consistently better on cold-start items compared to the strongest baseline CVAE, with significantly better performance on normal items. This further demonstrates the superior cold-start item recommendation ability of the proposed MDsym-CVAE.

\begin{figure}
\centering
\begin{subfigure}[b]{0.46\textwidth}
\includegraphics[width=\textwidth]{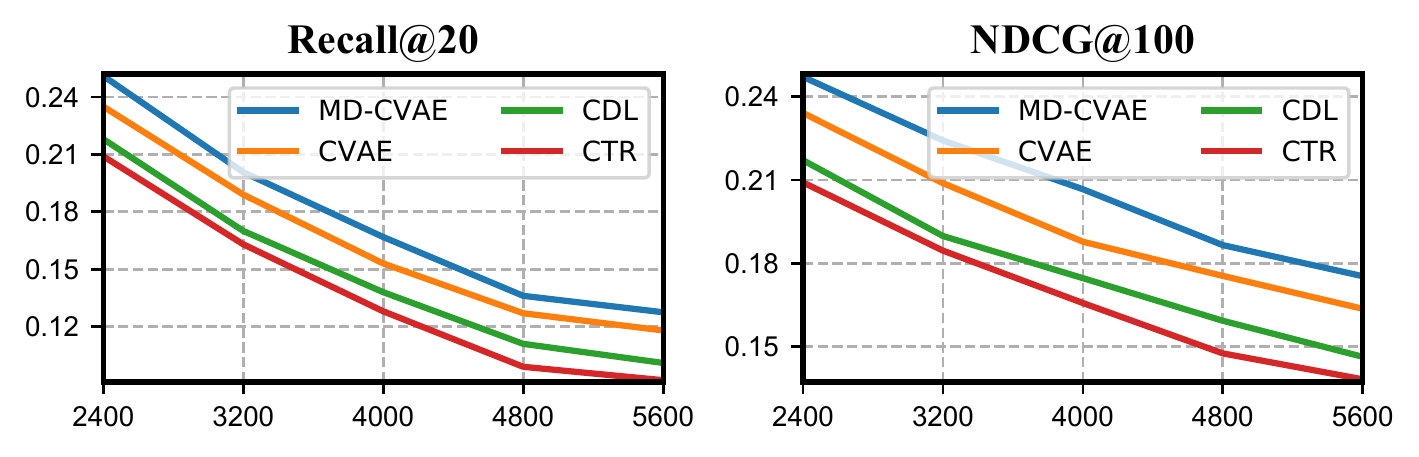}
\caption{\textit{citeulike-a} dataset}
\end{subfigure}

\begin{subfigure}[b]{0.46\textwidth}
\includegraphics[width=\textwidth]{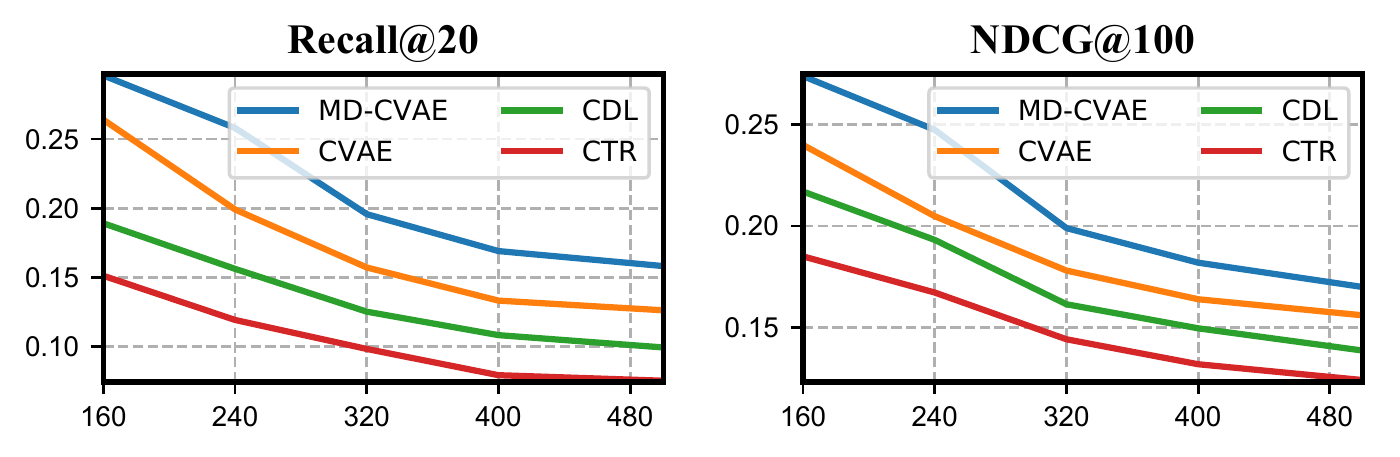}
\caption{\textit{movielen-sub} dataset}
\end{subfigure}

\begin{subfigure}[b]{0.46\textwidth}
\includegraphics[width=\textwidth]{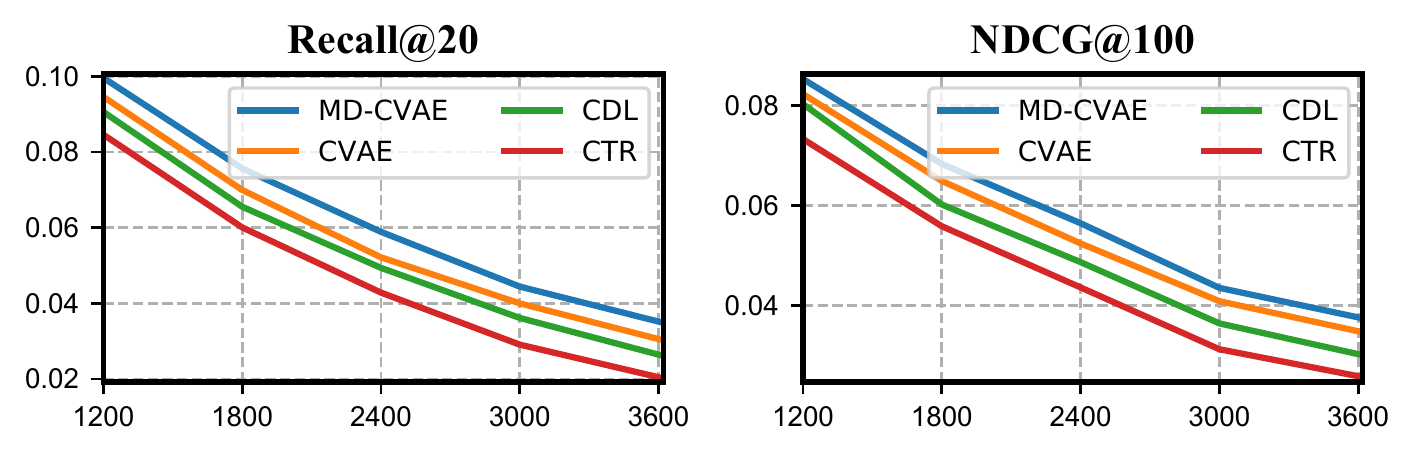}
\caption{\textit{toys \& games} dataset}
\end{subfigure}

\caption{The influence of the number of cold-start items on the recommendation performance for these items.}
\vspace{-3mm}
\label{fig:wrt_coldnum}
\end{figure}

Finally, we vary the number of cold-start items and evaluate the baselines and MDsym-CVAE. The results are illustrated in Fig. \ref{fig:wrt_coldnum}. From Fig. \ref{fig:wrt_coldnum} we can find that the recommendation performance on cold-start items drops as their number increases as expected. However, the decrease is slower for \textit{citeulike-a} dataset. The reason could be that item features in \textit{citeulike-a} dataset are extracted from the article title and abstract, which are directly related to the item content, whereas for \textit{movielen-sub} and \textit{toys \& games} datasets, the item features are extracted from noisier crawled plots and reviews. Therefore, better cold-start recommendation performance of MDsym-CVAE can be achieved if the extracted item content features are less noisy and more recommendation-oriented.

\section{Conclusions}

In this paper, we have proposed MD-CVAE to address two problems of vanilla UAEs in recommendations: sparsity and cold-start items. MD-CVAE seamlessly integrates latent item embeddings with UAE, allowing the incorporation of item content information to support recommendations in user-oriented models. Moreover, MD-CVAE can be easily generalized to multi-class classification tasks where new classes constantly appear after model training and deployment, such as automatic image censoring, \textit{etc.,} with a little adaptation. Therefore, we speculate that MD-CVAE could have a broader impact on tasks other than recommendations discussed in this paper.

\section*{Acknowledgment}
This research was supported in part by Tencent.

\bibliographystyle{ACM-Reference-Format}
\balance
\bibliography{vbae-cf}

\end{document}